\documentstyle[aps,twocolumn]{revtex}
\input epsf

\begin{document}


\draft

\title{Nonspherical perturbations of critical collapse and cosmic censorship}

\author{Carsten Gundlach}
\address{Max-Planck-Institut f\"ur Gravitationsphysik
(Albert-Einstein-Institut), Schlaatzweg 1, 14473 Potsdam, Germany}

\date{12 October 1997, revised 14 March 1998}

\maketitle


\begin{abstract}
  
Choptuik has demonstrated that naked singularities can arise in
gravitational collapse from smooth, asymptotically flat initial data,
and that such data have codimension one in spherical symmetry. Here we
show, for perfect fluid matter with equation of state $p=\rho/3$, by
perturbing around spherical symmetry, that such data have in fact
codimension one in the full phase space, at least in a neighborhood of
spherically symmetric data.

\end{abstract}



\subsubsection{Cosmic censorship and critical collapse}


Spacetime singularities are a ubiquitous feature of general
relativity. In astrophysical situations, singularities only arise
inside black holes. While no information (light signal) can leave the
black hole region of spacetime, the singularity itself cannot emit any
light signal at all: It is spacelike, that is, it can be thought of as
a local end to time. Nevertheless, the field equations of general
relativity also admit solutions with timelike singularities. Such a
singularity may be loosely thought of as a point in space, just like a
particle, and in many solutions light signals could travel from such a
singularity to infinity. The spacetime to the future of the
singularity is then no longer uniquely determined by initial data in
the past. It is therefore an interesting question to ask how generic
or natural such spacetimes, or the initial data from which they
evolve, are. A ``cosmic censorship conjecture'' has been formulated by
various authors, stating roughly that naked singularities do not arise
from asymptotically flat data for reasonable matter \cite{Wald}. The
only counterexamples existing until recently could be written off as
either not asymptotically flat, involving matter models (dust) that
form singularities even in the absence of gravity, or requiring high
symmetry (spherical shells).  Until recently, however, it could still
be conjectured that no smooth, asymptotically flat data for a
reasonable matter model would evolve into a naked singularity.  That
was disproved by the spacetimes constructed numerically by Choptuik
\cite{Choptuik}, which we review now.

Consider a smooth one-parameter family of spherically symmetric,
asymptotically flat, smooth initial data for a self-gravitating scalar
field. The only condition on the family is that it contains both data
sets which form a black hole (say, for large parameter $p$), and data
sets which do not (say, for small $p$). Choptuik showed that by
fine-tuning the value of $p$ to a critical value $p_*$, one can obtain
arbitrarily small black holes. For $p\gtrsim p_*$ the black hole mass
scales approximately as $M_{\rm BH}\simeq C (p-p_*)^\gamma$, with
$\gamma\simeq 0.37$ a universal constant.  A solution evolving from
fine-tuned initial data approaches one and the same solution,
independently of what the initial data looked like. The better the
fine-tuning, the longer the solution follows this ``critical
solution''. The critical solution is an attractor of codimension one,
because it has precisely one growing perturbation mode. Its basin of
attraction is the black hole threshold, because the eventual fate of
the solution depends on the amplitude of the growing mode in the
following manner. If it is present with, say, positive amplitude, the
final outcome will be a black hole. If it is present with negative
amplitude, the final outcome will be dispersion of the scalar waves to
infinity, without a black hole.  If its amplitude is precisely
zero, in the limit of infinitely good fine-tuning of the parameter
$p$, the solution settles down to the critical solution and never
leaves it.

The crucial point for cosmic censorship is that the critical solution
has a naked curvature singularity. Any smooth, asymptotically flat
initial data in which the growing mode is not present approach the
critical solution and its naked singularity. These data form a set of
codimension one (in spherical symmetry) because only the amplitude of
the one growing mode needs to be fine-tuned to zero. The spacetimes
arising from data on the black hole threshold do not contain a ``zero
mass black hole'', but contain a region of the universal critical
solution surrounding the naked singularity, which is smoothly matched
to an asymptotically flat region that depends on the initial
data. Near the singularity the deviation of the actual collapse
spacetime from the universal critical solution decays as a
(noninteger) power of geodesic distance to the singularity.  For a
recent review of critical collapse, see
\cite{critreview}.


\subsubsection{Beyond spherical symmetry}


An obvious question is if critical collapse is tied to spherical
symmetry. Abrahams and Evans \cite{AE} fine-tuned the collapse of
axisymmetric gravitational waves, and found evidence for black hole
mass scaling, universality, and a self-similar critical solution.
This shows that critical phenomena can occur for axisymmetric, highly
non-spherical initial data. In a complementary approach, we take here
a known critical solution in spherical symmetry, and perturb it
non-spherically. We find that the only growing mode is the previously
known spherical one. 

This result is weaker than that of Abrahams and Evans in being only
perturbative. But a linearized result establishes a result in an open
neighborhood, up to linearization stability. In spherical symmetry
the basin of attraction of the critical solution was empirically found
to be the entire black hole threshold, even far from the critical
solution \cite{Choptuik}. This makes nonlinear instabilities seem
extremely unlikely, and suggests that the open set may be in fact
quite large.  Generally it is remarkable how much in critical collapse
can be understood quantitatively through linear perturbation
calculations: not only the critical exponent $\gamma$, but also
universality with respect to initial data and matter models, and
critical exponents for the black hole electric charge and angular
momentum.

Our result is stronger than that of Abrahams and Evans in not being
limited to axisymmetry. Choptuik's spherically symmetric results
showed that a genericity condition on initial data is necessary in
formulating cosmic censorship. Here we have demonstrated for the first
time that this condition is sharp: The set of ``non-generic'' initial
data that form a naked singularity can locally have codimension one in
the full phase space.


\subsubsection{The background solution}


We write the general spherically symmetric spacetime as a manifold
$M=M^2 \times S^2$ with metric
\begin{equation}
g_{\mu\nu} = {\rm diag}\left(g_{AB}, r^2 \gamma_{ab}\right),
\end{equation}
where $g_{AB}$ is an arbitrary metric on $M^2$, $r$ is a scalar on
$M^2$, with $r=0$ defining the boundary of $M^2$, and $\gamma_{ab}$ is
the unit curvature metric on $S^2$. This spacetime is continously
self-similar if there are coordinates $x$ and $\tau$ on $M^2$ such
that in these coordinates the rescaled metric coefficients $e^{2\tau}
g_{AB}$ and $e^{2\tau} r^2$ do not depend on $\tau$. It is discretely
self-similar if they are periodic in $\tau$. A linearized spherical
perturbation of a continously self-similar spacetime can be decomposed
into modes of the schematic form $e^{\lambda\tau} f(x)$. The critical
solution has precisely one mode with $Re\lambda>0$. This is the mode
that has to be beaten down by fine-tuning, while all other modes die
away naturally as the curvature singularity $\tau=\infty$ is
approached.

As a matter model we choose the perfect fluid with equation of state
$p={c_s^2}\rho$, with ${c_s^2}$ a constant. While our equations hold
for $0<c_s^2<1$, we have carried out the numerical calculations only
for $c_s^2=1/3$, the equation of state of a radiation gas. 

The coordinates $x$ and $\tau$ adapted to self-similarity are not
unique, and they need not be fixed for the purpose of the following
discussion. In our numerical calculations, however, we make a
coordinate choice that is based on the Schwarzschild-like form
$ds^2=-{N}^2\,dt^2+{A}^2\,dr^2+r^2\,d\Omega^2$ of the metric.  The
coordinate transformation $t=-e^{-\tau}$, $r=sxe^{-\tau}$ brings this
into the form
\begin{eqnarray}
\label{bgmetric}
g_{AB} &&  = e^{-2\tau}\left(
\begin{array}{cc}
-{N}^2 + s^2x^2{A}^2 &  -s^2x{A}^2 \\
{\rm Symm} &  s^2{A}^2
\end{array}\right), \\
r^2 && = e^{-2\tau}s^2x^2.
\end{eqnarray}
As a final gauge condition, we impose ${N}=1$ at $x=0$, that is, $t$
is central proper time.  The spacetime in these coordinates is
self-similar if $N(x,\tau)$ and $A(x,\tau)$ are functions of $x$
only. 

The background stress-energy tensor is
\begin{equation}
t_{\mu\nu} = {\rm diag}\left(t_{AB},{c_s^2}\rho r^2 \gamma_{ab}\right),
\end{equation}
where
\begin{equation}
t_{AB} = (1+{c_s^2}) \rho u_A u_B + {c_s^2}\rho g_{AB},
\end{equation}
and $u_\mu=(u_A,0)$ is the fluid 4-velocity. The calculation of the
background solution as a non-linear boundary value problem was carried
out along the lines of
\cite{EC,KHA,Maison}. 
For the density we make the ansatz $\rho=e^{2\tau}\bar\rho(x)$, and
for the radial velocity, $u^Ar_{,A}=v(x)$. Under this ansatz the
Einstein and matter equations go over into a system of coupled
ordinary differential equations (ODEs). These have a regular singular
point (sonic point) where the surface $x={\rm const}$ becomes an
ingoing matter characteristic. The solution we want is the one
uniquely specified by regularity at the origin $x=0$ and at the sonic
point. We define the constant $s$ so that the sonic point is
at $x=1$, and solve the ODEs as a boundary value problem between $x=0$
and $x=1$, solving for $s$ as a nonlinear eigenvalue. 


\subsubsection{Perturbation method}


Now one could perturb the critical solution $Z(x)$ with an ansatz
$\Delta Z(x,\tau) = e^{\lambda\tau}f(x)$ and solve the linearized
boundary value problem for the discrete (complex) eigenvalues
$\lambda$ and mode functions $f(x)$. This program has been carried out
both for continously \cite{KHA,Maison} and discretely
\cite{Gundlach} self-similar solutions, and has
allowed a precise numerical calculation of the critical exponent
$\gamma$, which can be shown by dimensional analysis to be simply the
inverse of the one positive real eigenvalue $\lambda$. As a byproduct,
one can determine the entire perturbation spectrum (which is discrete).

Here, we mainly want to know is if there is {\it any} eigenvalue with
positive real part among the non-spherical perturbations, besides the
one in the spherical perturbations. To answer this yes/no question, we
write the perturbation equations as evolution equations in the time
variable $\tau$.  We decompose into spherical harmonics, and consider
each value of $l$ and $m$ separately.  Because the background is
spherically symmetric, the dynamics of perturbations are the same for
all values of $m$ (for given $l$). We evolve generic initial
data for these equations for a sufficiently large interval of
$\tau$. Generic data, with no field vanishing anywhere, constitute a
superposition of {\it all} the (unknown) perturbation modes.

In the time evolution, the mode with the largest $Re\lambda$ takes
over after a transition period, and both that $\lambda$ and the
corresponding $f(x)$ can be simply read off from the late-time data
\cite{KHA2}. For $l=0$ in particular, this allows us to check our
procedure by reading off the critical exponent as $\gamma\simeq 0.36$
in agreement with previous calculations \cite{KHA,Maison}. 

On a numerical grid, the frequency of modes that can be represented is
limited by the grid spacing, so that we are only probing a large
finite-dimensional subspace of all possible modes. Nevertheless, one
can rule out the existence of unstable modes at very high spatial
frequency (with respect to the coordinate $x$) by the following
argument. The perturbation equations form a system of linear wave
equations with $x$-dependent coefficients and an $x$-dependent mass
matrix. High-frequency modes propagate essentially by geometric
optics, and the mass terms are irrelevant for them. Therefore all
high-frequency modes have the same dynamics. If they are shown to be
decaying at frequencies still resolved by the numerical grid but at
which the mass-like terms can already be neglected with respect to
derivative terms, then we can be sure that even higher frequencies not
resolved on the numerical grid will decay in the same way.
The geometric optics argument also applies to large values of $l$,
which labels spatial frequencies in the angular coordinates $\theta$
and $\varphi$. However, high $l$ modes actually decay faster in $\tau$
with increasing $l$, by a factor of $e^{-l\tau}$. Roughly speaking,
this is because regular perturbations must be of $O(r^l)$ at the origin.

The spherical perturbation equations are obtained simply by
linearizing the spherical field equations. There is no difficulty in
writing them as evolution equations in time coordinate $\tau$ and
radial coordinate $x$.  Boundary conditions at $x=0$ arise from
demanding regularity at the center of spherical symmetry (all fields
must be either even or odd in $x$). The surface $x=1$ is an ingoing
characteristic of the spherical perturbation equations, that is, a
sound cone. No physical boundary conditions for perturbations are
required there. Numerically, this absence of boundary conditions is
implemented by a finite difference scheme that is aware of the
characteristics. Then we can bound the numerical domain by $x=0$ on
one side and $x=1$, or any larger constant $x$, on the other and
evolve to arbitrary values of $\tau$.  If one allows for non-spherical
perturbations, one encounters gravitational as well as sound
waves. The numerical domain for the entire system of perturbations
then has to be extended to the ingoing light cone, that is to $x=x_c$
defined by $sx_c{A}(x_c)={N}(x_c)$. The background solution is easily
extended from $x=1$ to $x=x_c$ and beyond. Fig. 1 illustrates the
coordinates, the characteristics, and the numerical domain.


\subsubsection{Gauge-invariant perturbations}


In going beyond spherical symmetry, one also has to deal with
gauge-dependence and the presence of constraints in the linearized
Einstein and matter equations. Throughout this letter, we use the
formalism and notation of Gerlach and Sengupta (GS)
\cite{GS}.
Any linear perturbation around spherical symmetry can be decomposed
into scalar, vector and tensor fields on $M^2$ times spherical
harmonics $Y_l^m$ on $S^2$. Different $l,m$ decouple. In the following
we consider one value of $l,m$ at a time, and no longer write these
indices on $Y$. Spherical harmonic vector fields on $S^2$ are $Y_{:a}$
and $S_a={\epsilon_a}^b Y_{:b}$, where a colon indicates the covariant
derivative on $S^2$, $\gamma_{ab:c}=0$, and $\epsilon_{ab}$ is the
covariantly constant totally antisymmetric tensor on $S^2$,
$\epsilon_{ab:c}=0$. Tensor perturbations based on $Y$ and its
derivatives (even or polar perturbations) decouple from tensor
perturbations based on $S_a$ and its derivatives (odd or axial
perturbations). In the following we consider even and odd
perturbations separately.

GS express the 10 metric perturbations in the $2+2$ split. They
calculate their transformation to linear order under the 4
infinitesimal coordinate changes, and find 6 linear combinations of
metric perturbations (with coefficients depending on the background)
that are gauge-invariant to this order. The 10 stress-energy
perturbations can be combined to form 10 gauge-invariant ones. The
linearized Einstein equations can then be expressed in terms of those
16 variables alone. The remaining 4 variables are pure gauge in the
sense that one can give them arbitrary values, and reconstruct all 20
metric and stress-energy perturbations in the particular gauge
determined that way.

Calculation with, and interpretation of, the GS variables is
simplified by the fact that there is a gauge in which they can be
identified directly with 16 gauge-dependent perturbations, while the
remaining 4 gauge-dependent metric perturbations vanish. This gauge is
the Regge-Wheeler gauge. In order to keep the notation simple, we present
the 16 gauge-invariant perturbations in this manner. The split between
the odd and even sectors is as follows. There are 3 odd metric
perturbations, and 1 odd infinitesimal coordinate transformation,
leaving 2 odd gauge-invariant perturbations, in the form of a vector
field on $M^2$. There are $7-3=4$ even gauge-invariant metric
perturbations, in the form of a symmetric tensor and a scalar. There
are 3 odd and 7 even gauge-invariant matter perturbations. The general
odd metric and matter perturbations, in Regge-Wheeler gauge but
expressed through the gauge-invariants, are
\begin{eqnarray}
\Delta g_{\mu\nu} && = \left(
\begin{array}{cc}
0 &  k_A Y_{:b} \\
{\rm Symm} &  0
\end{array}\right), 
\\
\Delta t_{\mu\nu} && = \left(
\begin{array}{cc}
0 &  L_A Y_{:b}\\
{\rm Symm} &  L(S_{a:b} + S_{b:a})
\end{array}\right).
\end{eqnarray}
The general even
metric and matter perturbations are
\begin{eqnarray}
\label{evenmetric}
\Delta g_{\mu\nu} && = \left(
\begin{array}{cc}
k_{AB} &  0 \\
0 &  kr^2 \gamma_{ab} 
\end{array}\right),
\\
\Delta t_{\mu\nu} && = \left(
\begin{array}{cc}
T_{AB} &  T_A Y_{:b}\\
{\rm Symm} &  T^1r^2 \gamma_{ab} + T^2 Y_{:ab} 
\end{array}\right).
\end{eqnarray}

We now define gauge-invariant fluid perturbations, and present them
once more by giving the equivalent gauge-dependent perturbations in
Regge-Wheeler gauge. The odd perturbation of the fluid 4-velocity is
\begin{equation}
\Delta u_{\mu}=(0,\beta S_{a}). 
\end{equation}
$\beta$ parameterizes axial fluid rotation.
The even perturbation of the fluid 4-velocity is 
\begin{equation}
\Delta u_{\mu}=(\Delta u_A, \alpha Y_{:a}),
\hbox{ where } \Delta u_A=\gamma n_A + k_{AB} u^B.
\end{equation}
Here $n^A = \epsilon^{AB} u_B$, with $\epsilon_{AB}$ the totally
antisymmetric covariant unit tensor on $M^2$. Note that $u^A\Delta
u_A=0$.  $\alpha$ parameterizes axial fluid motion, while $\gamma$
parameterizes perturbations of the radial fluid motion. The density
perturbation also belongs to the even sector, and we parameterize it
as $\Delta \rho =
\omega Y
\rho$. (By virtue of our equation of state, $\Delta p = {c_s^2} \Delta\rho$.)
The general gauge-invariant stress-energy perturbations of GS are
related to the gauge-invariant fluid perturbations as follows. In the
odd sector, we have
\begin{equation}
L_A = \beta (1+{c_s^2}) \rho u_A, \qquad L=0,
\end{equation}
and in the even sector,
\begin{eqnarray}
&& T_A = \alpha (1+{c_s^2}) \rho u_A, \quad T^1 = (\omega + k) {c_s^2} \rho,
\quad T^2 = 0, \\
&& T_{AB} = \omega t_{AB} + 2\Delta u_{(A} u_{B)} (1+{c_s^2})
\rho + k_{AB} {c_s^2}\rho.
\end{eqnarray}


\subsubsection{Odd perturbations}


We must now extract a well-posed initial value problem from the
gauge-invariant perturbation equations. For the odd sector this is
straightforward. The one nontrivial matter conservation equation is
$(r^2 L_A)^{|A}=0$ (where $g_{AB|C}= 0$), or
\begin{equation}
(\beta r^2 \rho u^A)_{|A} = 0.
\end{equation}
This equation is an advection equation for $\beta$, and can be solved
independently of all other perturbations, from $\beta$ given on an initial
spacelike hypersurface of $M^2$. GS have shown that by defining the
scalar $\Pi = \epsilon^{AB}(r^{-2}k_A)_{|B}$ the Einstein equations
for $k_A$ can always be reduced to the scalar wave equation
\begin{equation} 
[r^{-2}(r^4\Pi)_{|A}]^{|A}-(l+2)(l-1)\Pi=16\pi \epsilon^{AB}L_{A|B},
\end{equation}
which they call the odd-parity master equation. The full $k_A$ can be
reconstructed by quadratures once this equation has been solved for
$\Pi$. (Note that the source $L_A$ is already known in the case of
perfect fluid matter.) 


\subsubsection{Even perturbations}


The even perturbations are more entangled. For $\alpha$ we once again
have an advection equation, but now with sources. The first-order
equations for the density perturbation $\omega$ and radial velocity
perturbation $\gamma$ can be combined to form a single wave equation
at the speed of sound for either $\omega$ or $\gamma$. For fixed
metric perturbation $k_{AB}$ and $k$, the initial value problem would
then be clear. Unfortunately, one apparently cannot extract a master
equation for the metric perturbations from the linearized Einstein
equations for arbitrary matter, but one always has a system with
constraints. Instead, we follow Seidel \cite{Seidel} in first
focussing attention on those components of the linear Einstein
equations with vanishing matter sides. 4 such components exist,
because the 7 even stress-energy perturbations are linear in only the
3 even matter perturbations. One of these is
\begin{equation}
k_A^{\ A}=-16\pi T^2=0. 
\end{equation}
It is natural to decompose the trace-free tensor
$k_{AB}$ covariantly into two scalars, via
\begin{equation}
k_{AB} = \phi(u_A u_B + n_A n_B) + \psi (u_A n_B + n_A u_B),
\end{equation} 
and to introduce the frame derivatives
\begin{equation}
\label{dotprime}
\dot f = u^A f_{|A}, \qquad f' = n^A f_{|A}.
\end{equation}
The three remaining source-free Einstein equations can
then be written as
\begin{eqnarray}
\label{evenEOM}
 - (\dot \chi)\dot{} + (\chi')' && =  S_\chi, \\
 - (\dot k)\dot{} +  {c_s^2} (k')'  && =  S_k, \\
 -\dot \psi  && = S_\psi.
\end{eqnarray}
where $\chi=\phi - k$ replaces $\phi$. The source terms $S_\chi$,
$S_k$ and $S_\psi$ are linear in $\chi$, $k$, $\psi$ and their first
derivatives $\chi'$, $k'$, $\psi'$, $\dot\chi$, and $\dot k$, but do
not contain $\dot \psi$. While the highest derivatives of $\chi$ form
a wave equation with characteristics given by the metric $g_{AB}=-u_A
u_B + n_A n_B$, $k$ obeys a wave equation with characteristics given
by the ``fluid metric'' $-  u_A u_B + {c_s^{-2}}n_A n_B$. These
characteristics have speed $c_s$ relative to the
fluid. Finally, $\psi$ is advected with the fluid. Therefore $\chi$
characterizes gravitational waves, $k$ sound waves, and $\psi$ polar
fluid flow. The metric perturbation $k$ ``knows about'' the speed of
sound because we have used $\Delta p = dp/d\rho \Delta\rho = c_s^2
\Delta\rho$ in finding the source-free linearized Einstein equations.
If we add to these three equations the identity $(f'u^A)_{|A}=(\dot f
n^A)_{|A}$ and the definitions (\ref{dotprime}), we have a complete
first-order system of equations. The variables $\chi$, $\dot\chi$,
$k$, $\dot k$ and $\psi$ can be set freely on a spacelike hypersurface
in $M^2$. From the equations of motion one can see that any regular
solution must scale at $r=0$ as $k\sim r^l$, $\psi \sim r^{l+1}$, and
$\chi \sim r^{l+2}$. This follows also from the requirement that the
metric perturbation (\ref{evenmetric}) be a regular tensor in four
dimensions at $r=0$. Furthermore, the perturbed metric remains
continuously self-similar if $k$, $\psi$ and $\chi$ are independent of
$\tau$. The perturbed metric is discretely self-similar if these
fields are periodic in $\tau$.

The Einstein equations we have not used yet give the matter
perturbations directly in terms of the metric perturbations. As a
check of the correctness of our equations and their numerical
implementation, we have numerically differentiated the numerical
solution and verified that the perturbed matter equations of motion
(or Bianchi identities) are obeyed.

The case of $l=1$ even perturbations is not covered by the framework
of GS and has to be treated separately. Clearly $l=1$ does not admit
gravitational waves, so that we can use Newtonian intuition. The
$l=1$ matter perturbations are pure gauge, corresponding to an initial
displacement and velocity of the spherical background solution.


\subsubsection{Numerical method}


We conclude with a remark on the numerical implementation. The
linearized field equations are of the form
\begin{equation}
{\partial u\over \partial \tau} = A {\partial u \over \partial x} + B
u.
\end{equation}
In order to make the numerical evolution stable even
though the lines of constant $x$ change from timelike to spacelike in
our computational domain, it is essential to use a characteristic
scheme \cite{Leveque}. Let $V$ be the matrix of (column) eigenvectors
of $A$. Let $\Lambda$ be the diagonal matrix composed of the
corresponding eigenvalues. Then $A=V \Lambda V^{-1}$. Let $\Lambda_+$
be the diagonal matrix with zeros in the place of the negative
eigenvalues. Define $\Lambda_-$, $A_+$ and $A_-$ in the obvious
manner, so that $A=A_++A_-$. We have used the numerical
scheme 
\begin{equation}
{u^{n+1}_j - u^n_j \over \Delta \tau}
= A_+ {u^n_{j+1} - u^n_j \over \Delta x} +  
A_- {u^n_j - u^n_{j-1} \over \Delta x} + B u^n_j,
\end{equation}
which is first-order accurate, and stable even for superluminal shift.
The matrix $A$ is just sparse enough for its eigenvalues and
eigenvectors to be calculated in closed form. As expected, the
characteristics are the fluid world lines, light cones and sound
cones.


Helpful conversations with M. Alcubierre, W. Junker, B. Schmidt and
E. Seidel, and a communication from U. Gerlach, are gratefully
acknowledged.



\begin{figure}
\epsfxsize=8cm
\epsffile{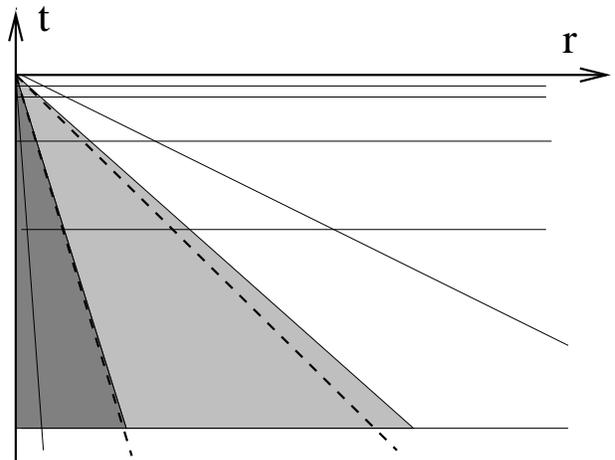}
\caption{The horizontal thin lines accumulating at
$t=0$ are lines of constant $\tau$. The fanning thin lines are lines
of constant $x$. The inner dashed line is the ingoing sound
cone. The outer dashed line is the ingoing light cone.  The
dark shaded region is the numerical domain of the background,
spherical and odd perturbations calculations. The total shaded region
is the numerical domain of the even perturbations calculation.}
\label{convergence}
\end{figure}


\end{document}